\documentclass[aps,prl,floatfix,twocolumn,preprintnumbers,amsmath,amssymb,groupedaddress]{revtex4-1}

\usepackage{graphicx}  
\usepackage{dcolumn}   
\usepackage{bm}        
\usepackage{color}     
\usepackage{amsmath,amssymb}
\usepackage{psfrag}

\newcommand{\ua}{\uparrow}
\newcommand{\da}{\downarrow}

\begin{document}

\title{Spin relaxation mechanism in graphene: resonant scattering by magnetic impurities}
\author{Denis Kochan, Martin Gmitra, and Jaroslav Fabian}
\affiliation{Institute for Theoretical Physics, University of Regensburg, 93040 Regensburg, Germany\\
}
\begin{abstract}
It is proposed that the observed small (100~ps) spin relaxation time in graphene is due to resonant scattering by local magnetic moments.
At resonances, magnetic moments behave as spin hot spots: the spin-flip scattering rates are as large as the spin-conserving ones,
as long as the exchange interaction is greater than the resonance width. Smearing of the resonance peaks by the presence of
electron-hole puddles gives quantitative agreement with experiment, for about 1~ppm of local moments.
While the local moments can come from a variety of sources, we specifically focus on hydrogen adatoms.
We perform first-principles supercell calculations and introduce an effective Hamiltonian to obtain
realistic input parameters for our mechanism.
\end{abstract}


\keywords{spin relaxation, graphene, magnetic moments}
\date{\today}
\maketitle

Graphene~\cite{Geim07, Neto11} has been considered an ideal
spintronics \cite{Zutic2004:RMP, Fabian2007:APS} material. Its spin-orbit coupling
being weak, the spin lifetimes of Dirac electrons are expected to be long, on
the order of microseconds~\cite{Pesin2012:NM}. Yet experiments find tenths of a nanosecond
~\cite{Tombros2007:N,Tombros08,Pi2010:PRL,Yang2011,Han2011,Avsar2011,Jo11,Mani2012:NC}.
This vast discrepancy has been the most outstanding puzzle of graphene
spintronics. Despite intense theoretical efforts
\cite{Huertas2006,Dora2010:EPL,Jeong2011:PRB,Dugaev2011:PRB,
Ertler2009:PRB,Zhang2011:PRB,Ochoa2012:PRL,Fedorov2013:PRL},
the mechanism for the spin
relaxation in graphene has remained elusive.
Recently, mesoscopic transport experiments \cite{Lundeberg2012} found evidence that local magnetic moments
could be the culprits. Here we propose a mechanism of how even a small concentration of such
moments can drastically reduce the spin lifetime of Dirac electrons.
If the local moments sit at resonant scatterers, such as
vacancies \cite{Ugeda2010:PRL, Nair2012:NP, McCreary2012:PRL} and adatoms \cite{Yazyev2010:RPP, McCreary2012:PRL},
they can act as spin hot spots \cite{Fabian1998:PRL}: while contributing little to
momentum relaxation, they can dominate spin relaxation.
Although our mechanism is general, we specifically assume that local moments
come from hydrogen adatoms. The calculated spin relaxation rates for
1~ppm of local moments, when averaged over electron density fluctuations
due to electron-hole puddles, are in quantitative agreement with experiment.
Our theory shows that in order to increase the spin lifetime in graphene,
local magnetic moments at resonant scatterers need to be chemically
isolated or otherwise eliminated.

Magnetic impurities typically do not play a role in the electron spin relaxation of
conductors \cite{Zutic2004:RMP}, unless when doped with transition
metal elements. In graphene the presence of local magnetic moments is not
obvious, unless the magnetic sites (vacancies or adatoms) are intentionally
produced \cite{Nair2012:NP, McCreary2012:PRL}. It is reasonable to expect
that there are not more magnetic sites than, say, 1~ppm, in graphene samples
investigated for spin relaxation. For this concentration a simple estimate
gives a weak spin relaxation rate, similar to what is predicted for spin-orbit
coupling mechanisms. Indeed, the Fermi golden rule gives,
for exchange coupling $J$ between electrons and local moments,
spin relaxation rate $1/\tau_s \approx \tfrac{2\pi}{\hbar}\eta J^2 \nu_0(E_F)$, where
$\nu_0(E_F)$ is graphene's density of states at the Fermi level
and $\eta$ is the concentration of the moments.
Taking representative values of $J \approx 0.4$~eV, $\eta \approx 10^{-6}$,
and $E_F \approx 0.1$~eV (for which $\nu_0$ is about 0.01 states
per eV and atom), one gets spin relaxation times of 100~ns, three orders
below the experimental 100~ps.

What we show in this paper is that the
spin relaxation due to magnetic impurities in graphene is significantly
enhanced by resonant scattering, for which the perturbative
Fermi golden rule does not apply. The intuitive idea is that
if the exchange coupling $J$ is greater than the resonance width $\Gamma$,
the electron spin can precess at resonance by at least one period during
the interaction time with the impurity. Then the spin-flip probability
becomes as likely as the spin-conserving one. This idea is confirmed
by an explicit calculation for graphene with a hydrogen adatom. We
use a supercell first-principles band structure to investigate the local
magnetic moments (as has been done earlier \cite{Yazyev2010:RPP})  and parameterize the band structure in terms of
effective exchange couplings, to obtain their realistic estimates.
We then use a single impurity spin model with exchange on the
resonance site to calculate the T-matrix and spin relaxation rate.
Finally, we illustrate the intuitive picture of resonant spin enhancement
on a toy one-dimensional model of an electron scattering off a magnetic
moment in a resonant quantum well.

\begin{figure*}
\centering
\includegraphics[width=1.98\columnwidth,angle=0]{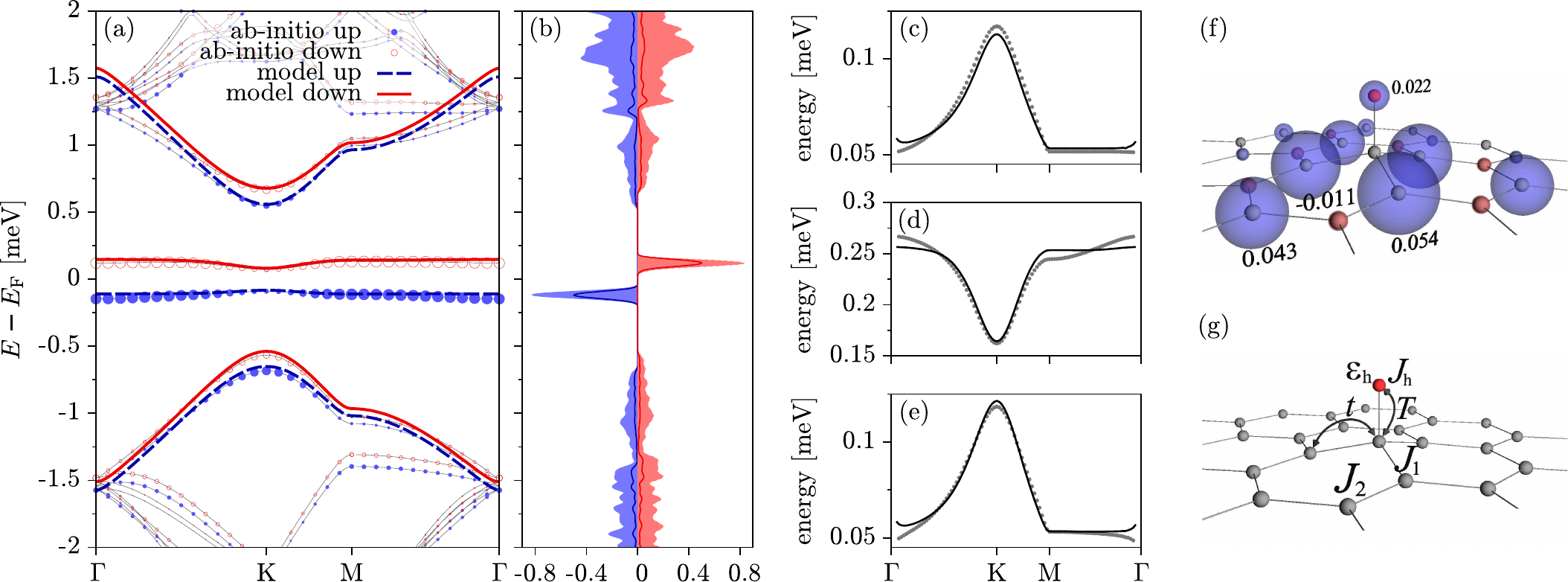}
\caption{First-principles results for a $5\times 5$ graphene
supercell with a single hydrogen adatom.
(a)~Spin-polarized band structure. The circles radii indicate the
presence of $p_z$ orbitals from the nearest neighbors to $\mathrm{C_H}$.
Bold lines (dashed and solid) come from the exchange hopping model, Eq.~(\ref{Eq:H'}).
(b)~Total density of states per atom (filled) and $p_z$ projected local
densities summed up to the third nearest carbon atoms to $\mathrm{C_H}$, normalized to the corresponding number of atoms in the set.
Exchange splittings of the conduction (c), mid-gap (d), and valence (e) bands. Solid lines are from the model.
(f)~Local magnetic moments around hydrogen, indicated in $\mu_B$.
(g)~Exchange hopping model of Eq.~(\ref{Eq:H'}).
}
\label{Fig:1}
\end{figure*}

\paragraph{Spin-polarized band structure of hydrogenated graphene.}
The electronic structure of a relaxed $5\times 5$ supercell with a single H atom on top of a C atom
(denoted below as $\mathrm{C_H}$) has been calculated within density functional theory using the full-potential linearized augmented plane wave
method as implemented in the FLEUR code \cite{fleur}.
Figure~\ref{Fig:1}(a-f) shows the results. The valence and conduction bands are separated at $\mathrm{K}$ point due
to covalent bonding of carbon $p_z$ and hydrogen $s$ orbitals by about 1~eV. In between lies
the mid-gap band formed mainly by $p_z$ orbitals of C atoms closest to H.
The ground state is ferromagnetic, with the exchange splitting
of about 0.1~eV. The magnetic moment is significant in a close neighborhood of $\mathrm{C_H}$ only,
oscillating as a function of position. The largest moment, of $0.054~\mu_\mathrm{B}$, is
on the nearest neighbors to $\mathrm{C_H}$. Hence the spin splitting of the mid-gap states
is maximal and gradually decreases for the bands away from the Fermi level, whose character is less
influenced by the H region.

To parameterize the first-principles data, we have extended the hopping Hamiltonian
studied in Refs.~\cite{Robinson2008:PRL,Wehling2010:PRL,Gmitra2013:PRL}. The scheme is in Fig.~\ref{Fig:1}(g).
The orbital effects due to H are captured by on-site
energy $\varepsilon_\mathrm{h}$ and hopping $T$. To this we add model exchange couplings
$J_\mathrm{h}$, $J_1$ and $J_2$, on the sites of large magnetic moments, inspired
by Fig.~\ref{Fig:1}(f). The impurity Hamiltonian $H'_{\rm eff}$, which is added to graphene's
$H_0=-t\sum_{\langle m,n\rangle} c^\dagger_{m} c^{\phantom\dagger}_{n}$
($t=2.6\,\mathrm{eV}$) is
\begin{equation}\label{Eq:H'}
\begin{aligned}
&H'_{\rm eff}=
\sum\limits_{\sigma} h^\dagger_\sigma (\varepsilon_\mathrm{h}-J_\mathrm{h}\hat{\sigma}_z)h^{\phantom{\dagger}}_\sigma
+
T(h^\dagger_\sigma c^{\phantom{\dagger}}_{\mathrm{C_H},\sigma}+c^\dagger_{\mathrm{C_H},\sigma} h^{\phantom{\dagger}}_\sigma)\\
&-
J_1\sum\limits_{m_{1},\sigma} c^\dagger_{m_{1},\sigma}\hat{\sigma}_z c^{\phantom\dagger}_{m_{1},\sigma}
-
J_2\sum\limits_{m_{2},\sigma} c^\dagger_{m_{2},\sigma}\hat{\sigma}_z c^{\phantom\dagger}_{m_{2},\sigma}\,.
\end{aligned}
\end{equation}
Here $h^\dagger_\sigma$~($h^{\phantom{\dagger}}_\sigma$) and $c^\dagger_\sigma$~($c^{\phantom{\dagger}}_\sigma$)
are fermionic creation (annihilation) operators acting on the hydrogen and graphene carbon sites,
respectively. Subscript $\sigma=\{\ua,\da\}$ stands for the spin component along the
$z$-direction (quantization axis); $\hat{\sigma}_z$ is the Pauli matrix.
Subscripts $m_1$ and $m_2$ label the three first-nearest and the
six second-nearest neighbors of $\mathrm{C_H}$.

Orbital parameters $\varepsilon_\mathrm{h}=0.16\,\mathrm{eV}$ and $T=7.5\,\mathrm{eV}$ were
fitted already in Ref.~\cite{Gmitra2013:PRL}. Least-square fitting the model Hamiltonian $H'_{\rm eff}$, Eq.~(\ref{Eq:H'}),
to our supercell spin-polarized first-principles data, gives $J_\mathrm{h}=-0.82\,\mathrm{eV}$, $J_1=0.69\,\mathrm{eV}$, and $J_2=-0.18\,\mathrm{eV}$. We fitted the valence, mid-gap, and conduction bands
at 100 equidistant points along $\mathrm{\Gamma K M \Gamma}$.
The fits, shown in Fig.~\ref{Fig:1}(a) and detailed in Fig.~\ref{Fig:1}(c-e), are remarkably good especially around K.
We find that $J_\mathrm{h}$ alone controls the exchange splitting of the valence and conduction
bands in a large region around $\mathrm{K}$ point.

\paragraph{Resonant scattering by magnetic impurities.}
To solve the magnetic scattering problem using $H'_{\rm eff}$ in the
single impurity limit is numerically demanding. However, the most important spin-flip
contribution is expected to come from the exchange coupling on the resonant scatterer (H atom) site \cite{SM}.
We thus neglect $J_1$ and $J_2$ terms and propose the reduced Hamiltonian,
$H'({\hat{\mathbf{S}}})$:
%
\begin{equation}\label{Eq:H'(S)}
H'({\hat{\mathbf{S}}})=\sum\limits_{\sigma}
\varepsilon_\mathrm{h} h^\dagger_\sigma h^{\phantom{\dagger}}_\sigma+
T(h^\dagger_\sigma c_{\mathrm{C_H},\sigma}^{\phantom{\dagger}}+c_{\mathrm{C_H},\sigma}^\dagger h^{\phantom{\dagger}}_\sigma)-
J\,\hat{\mathbf{s}}\cdot\hat{\mathbf{S}}\,.
\end{equation}
The exchange term describes the interaction of electron spin $\hat{\mathbf{s}}=h^{\dagger}_{\alpha}\hat{\boldsymbol{\sigma}}^{\phantom{\dagger}}_{\alpha\beta}h^{\phantom{\dagger}}_{\beta}$
and impurity moment $\hat{\mathbf{S}}$. We keep orbital parameters $\varepsilon_\mathrm{h}=0.16\,\mathrm{eV}$ and $T=7.5\,\mathrm{eV}$, and take a generic value $J=-0.4\,\mathrm{eV}$ for exchange. The spin relaxation rates, when
broadened by puddles, are hardly influenced by the precise value and the sign of $J$ \cite{SM}.

In the independent electron-impurity picture (we do not discuss  Kondo physics), total Hamiltonian
$H_0+H'({\hat{\mathbf{S}}})$ diagonalizes in the singlet ($\ell=0$) and triplet ($\ell=1$) basis
$|\ell,m_\ell\rangle$ (here $m_\ell$ runs from $-\ell$ to $\ell$).
Eliminating by downfolding (L\"{o}wdin transformation) H orbitals,
we arrive at the single-site impurity Hamiltonian,
\begin{equation}\label{Eq:H'(S)-fold}
H'_{\mathrm{fold}}({\hat{\mathbf{S}}})=\sum\limits_{\ell,m_\ell}\alpha_\ell(E)\,c^\dagger_{\mathrm{C_H},\ell,m_\ell}c^{\phantom{\dagger}}_{\mathrm{C_H}, \ell,m_\ell} \,,
\end{equation}
where the energy-dependent on-site coupling is,
\begin{equation}\label{Eq:alpha}
\alpha_\ell(E)=\frac{T^2}{E-\varepsilon_\mathrm{h}+(4\ell-3)J} \,,
\end{equation}
different for singlet and triplet states.

The T-matrix elements for the above impurity problem can be calculated as
(see, e. g.,~\cite{Hewson1993:CUP})
\begin{equation}\label{Eq:T(E)}
\mathrm{T}(E)_{\boldsymbol{\kappa}',\ell',m_{\ell'}|\boldsymbol{\kappa},\ell,m_\ell}=
\frac{1}{N_\mathrm{C}}\,\frac{\delta_{\ell,\ell'}\,\delta_{m_\ell,m_{\ell'}}\,\alpha_\ell(E)}{1-\alpha_\ell(E)G_0(E)} \,.
\end{equation}
where $\boldsymbol{\kappa}$ labels momentum and band index of graphene's Bloch states,
$N_\mathrm{C}$ is the number of carbon sites in the sample, and $G_0(E)$ is the retarded
Green function per carbon atom and spin of unperturbed graphene. Near the neutrality point ($E=0$),
$G_0(E)\simeq\tfrac{E}{D^2}\bigl[\ln{\bigl|\tfrac{E^2}{D^2-E^2}\bigr|}-i\pi\,\mathrm{sgn}(E)\Theta(D-|E|)\bigr]$,
where the graphene bandwidth $D=\sqrt{\sqrt{3}\pi}t\approx 6\,\mathrm{eV}$.

Resonant states appear for energies $|E|<D$ at which
the real part of the denominator of Eq.~(\ref{Eq:T(E)}) equals zero. Near the neutrality point
($|E|<<D$) we get the equation
\begin{equation}\label{Eq:E_res}
E_{\mathrm{res},\ell}\Bigl(1-\frac{T^2}{D^2}\ln\frac{E^2_{\mathrm{res},\ell}}{D^2}-\frac{T^2}{D^4}E_{\mathrm{res},\ell}^2\Bigr)=\varepsilon_\mathrm{h}-(4\ell-3)J
\end{equation}
which determines the resonant energies $E_{\mathrm{res},\ell}$ for singlet and triplet states.
For a non-magnetic impurity ($J=0$) there appears a single resonant level close
to the neutrality point \cite{Wehling2010:PRL}. For a magnetic impurity
this level splits to singlet and triplet peaks, and shifts in energy.
For $J<0$ the singlet resonance has a lower energy, see \cite{SM}.

From the T-matrix we obtain spin-flip rate $1/\tau_s$ at zero temperature (thermal broadening is discussed in \cite{SM}),
\begin{equation}\label{Eq:W_{s's}}
\begin{aligned}
1/\tau_s
=\eta \frac{2\pi}{\hbar}\nu_0(E)f_{-\sigma,\sigma}\Bigl(\tfrac{\alpha_1(E)}{1-\alpha_1G_0(E)},\tfrac{\alpha_0(E)}{1-\alpha_0G_0(E)}\Bigr),
\end{aligned}
\end{equation}
for the fraction of $\eta=N_\mathrm{A}/N_\mathrm{C}$ of impurities per carbon atom. Couplings $\alpha_\ell(E)$
are given by Eq.~(\ref{Eq:alpha}), $G_0(E)$ and $\nu_0(E)$ are graphene's Green function and DOS per atom and spin,
and auxiliary function $f_{\sigma,\sigma'}(x,y)$ is,
\begin{equation}\label{Eq:f}
f_{\sigma,\sigma'}(x,y)=
\frac{1}{2}\delta_{\sigma,\sigma'}\bigl|x\bigr|^2+
\frac{1}{8}\bigl|x+(\sigma\cdot\sigma')\,y\bigr|^2 \,.
\end{equation}
The spin-flip rate $1/\tau_s$ is peaked at resonances where
denominators $1-\alpha_\ell(E)G_0(E)$ have minima.

\begin{figure}
\centering
\includegraphics[width=0.98\columnwidth]{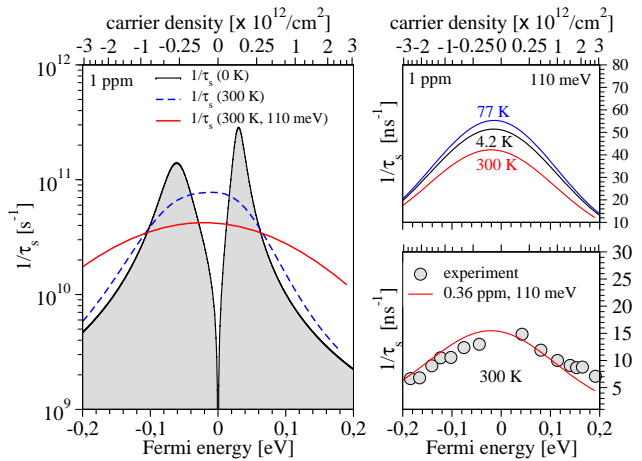}
\caption{Resonant enhancement of spin relaxation in graphene. Exchange $J = -0.4$~eV
and impurity fraction $\eta$ is indicated.
(a) Spin relaxation rate $1/\tau_s$ as a function of energy/carrier density, at 0~K, at 300~K, and at 300~K broadened
by puddles with energy fluctuations of 110~meV. (b) Broadened $1/\tau_s$ at different $T$.
(c) Comparison between theory and experiment (graphene data from Ref. \cite{Wojtaszek2013:PRB})
 at 300~K. }
\label{Fig:2}
\end{figure}

Spin relaxation rate $1/\tau_s$ is plotted in Fig.~\ref{Fig:2}, which is the main result of this paper.
Zero temperature rate shows singlet and triplet split resonance peaks, with widths $\Gamma$ of about 20 and 40~meV,
respectively. At 300~K the peaks merge. In realistic samples the neutrality point fluctuates
due to electron-hole puddles \cite{Hwang2007:PRL,Deshpande2009:PRB}. Also, different
magnetic impurities would give different peak positions and widths, providing additional
broadening. All such effects are modeled by gaussian energy broadening
with standard deviation $\sigma_{\mathrm{br}}$. In Fig.~\ref{Fig:2} we use
 $\sigma_{\mathrm{br}} = 110$~meV. From Fig.~\ref{Fig:2}(b) we can conclude
that the temperature dependence of $1/\tau_s$ is rather weak, essentially given by
Fermi broadening of the resonance structure. Finally, in Fig.~\ref{Fig:2}(c)
we compare the calculated spin relaxation rates with experiment, with adjusted
$\eta$. The agreement is remarkable. In fact, one can find a nice agreement for
a large window of $J$ (see \cite{SM}) by adjusting $\sigma_{\mathrm{br}}$ and $\eta$.
Vacancies and different adatoms are well covered by this mechanism.

In \cite{SM} we plot $1/\tau_s$ for ferromagnetic $J =0.4$~eV. The only effect,
after broadening, is the opposite (slight) skewness of the energy dependence (keeping
$\varepsilon_\mathrm{h}$ unchanged), coming from the flipped positions of the singlet
and triplet peaks. Also, in \cite{SM} we demonstrate that resonance enhancement of
$1/\tau_s$ is present for even much smaller $J$, as long as $J \agt \Gamma$,
confirming the intuitive picture of the enhancement coming from the spin precession
being faster than the leakage rate. One important conclusion one can draw from this
concerns spin-orbit coupling (SOC). Hydrogen adatoms induce SOC of about 1~meV
\cite{Balakrishnan2013:NP, Gmitra2013:PRL}. This is smaller than $\Gamma$, so the resonant enhancement
will be much less pronounced, unless $\eta$ is increased to,
say $10^{-3}$ \cite{SM}. Nevertheless, there could be heavier adatoms that induce both large spin-orbit coupling and resonant scattering so that resonance enhancement could be present. It was recently shown
that Si adatoms sitting on top of the carbon bonds could also give 100~ps spin-flip
times \cite{Fedorov2013:PRL}, but for concentrations of $\eta \sim 10^{-3}$, three orders
more than what is needed for magnetic resonant scatterers. It is possible that the mechanism is indeed resonance enhancement
of the spin-flip rates.
In fact, resonant scattering by spin-orbit coupling inducing impurities was already invoked to explain strong spin-flip scattering in alkali \cite{Mahanti1971:PL} and noble \cite{Phivos} metals.

There have already been spin relaxation experiments with hydrogenated graphene. According to our
theory, an $sp^3$ bonded hydrogen should increase the spin relaxation rate. Unfortunately, the experimental
results differ. In Ref.~\cite{Wojtaszek2013:PRB} the spin relaxation rate decreased upon hydrogenation. In
Ref.~\cite{Balakrishnan2013:NP} spin relaxation has not changed much, while in Ref.~\cite{McCreary2012:PRL} evidence for magnetic moments was provided based on a different model,
that of fluctuating magnetic fields. It is likely that the experimental outcomes depend on the hydrogenation
method. At present it is not possible to form a unique experimental picture with which we could gauge
our theory. But we stress that we use hydrogen only as a convenient model to formulate our mechanism
quantitatively. The Hamiltonian we use is rather generic, and the results are very robust as far as the details
in $J$ and other parameters are concerned. It is even possible that hydrogenation isolates existing
magnetic moments at vacancies, thereby increasing $\tau_s$, as seen in Ref.~\cite{Wojtaszek2013:PRB}.

\begin{figure}
\centering
\includegraphics[width=0.95\columnwidth]{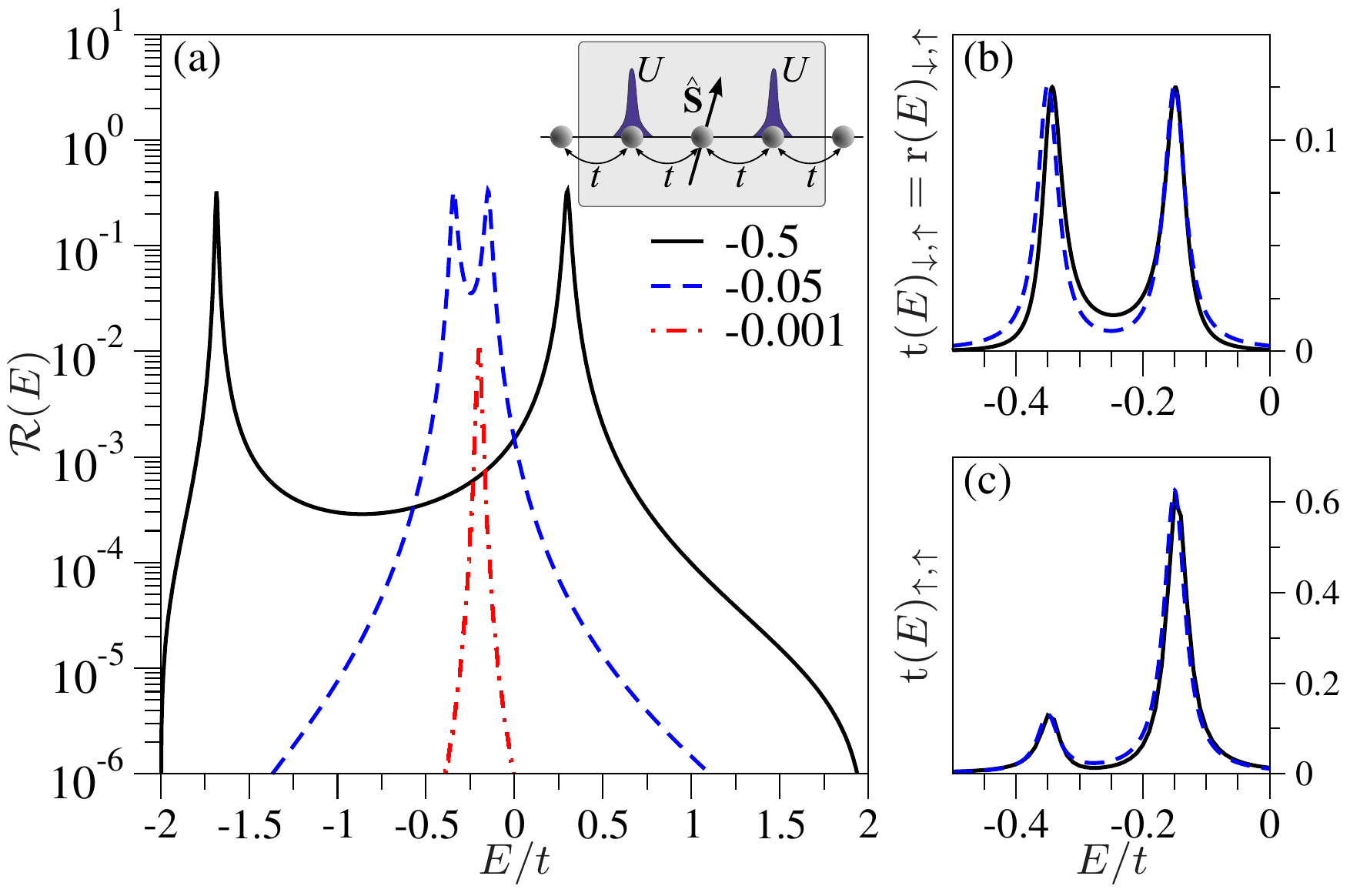}
\caption{Resonant enhancement of spin flips in
a one dimensional atomic chain with a double barrier hosting an impurity
spin. (a)~Ratio $\mathcal{R}(E)$ of spin-flip and spin-conserving transition probabilities
for $U/t=10$ and indicated $J/t$. Inset shows the model.
(b) Spin-flip $\mathrm{t}(E)_{\da,\ua}$ and (c) spin-conserving
$\mathrm{t}(E)_{\ua,\ua}$ probabilities for $J/t=-0.05$. The solid lines are exact
formulas, Eqs.~(\ref{Eq:t})-(\ref{Eq:r}), the dashed lines are
approximations, Eq.~(\ref{Eq:r-aprox}).
}
\label{Fig:3}
\end{figure}

\paragraph{Resonant spin-flip scattering in a one-dimensional double-barrier atomic chain.}
To make the resonant enhancement of the spin relaxation rate more transparent, we introduce
a toy model that captures all the essential features.
Consider an atomic chain with lattice constant $b$, whose central site ($m= 0$), trapped within
two $\delta$ barriers on its nearest neighbors, hosts the exchange interaction
$J\,\hat{\mathbf{s}}\cdot\hat{\mathbf{S}}$. The hopping Hamiltonian is
\begin{equation}\label{Eq:H_1D}
H=-t\sum\limits_{\langle m,n\rangle}(c^\dagger_mc^{\phantom{\dagger}}_n+ c^\dagger_n c^{\phantom{\dagger}}_m)+
U\sum\limits_{m=\mp 1}c^\dagger_{m}c^{\phantom{\dagger}}_{m}
-J\,\hat{\mathbf{s}}\cdot\hat{\mathbf{S}},
\end{equation}
as sketched in the inset of Fig.~\ref{Fig:3}(a).
In the singlet-triplet basis the transmission and reflection
amplitudes $\gamma_{\ell,m_\ell}(k)$ and $\beta_{\ell,m_\ell}(k)$,
are obtained analytically as
\begin{align}
\gamma_{\ell,m_\ell}(k)&=\frac{2it\bigl(1+U e^{ikb}/t\bigr)^{-1}\sin{kb}}{\bigl[E_k+J(4\ell-3)\bigr]\bigl(1+U e^{ikb}/t\bigr)+2te^{ikb}}\label{Eq:gamma}\,,\\
\beta_{\ell,m_\ell}(k)&=\gamma_{\ell,m_\ell}(k)-\frac{t+U e^{-ikb}}{t+U e^{ikb\phantom{-}}}\label{Eq:beta}\,.
\end{align}
The energy of the incident electron of momentum $k$ is $E_k=-2t\cos ({kb})$, and
the composite (electron and impurity) spin state $|\ell,m_\ell\rangle$, with angular momentum $\ell=1$ for triplet and $\ell=0$ for singlet
states; $m_\ell$ is the corresponding angular momentum projection (this index is
dropped in what follows, as neither amplitude depends on it).
We are interested in the transmission
$\mathrm{t}=|\gamma|^2$ and reflection $\mathrm{r}=|\beta|^2$ probabilities of various
spin transition processes $\sigma\rightarrow\sigma'$ so we trace out the impurity spin.
The result is
\begin{align}
\mathrm{t}(E_k)_{\sigma,\sigma'}&=f_{\sigma,\sigma'}\bigl(\gamma_1(k),\gamma_0(k)\bigr)\label{Eq:t}\,,\\
\mathrm{r}(E_k)_{\sigma,\sigma'}&=f_{\sigma,\sigma'}\bigl(\beta_1(k),\beta_0(k)\bigr)\label{Eq:r}\,,
\end{align}
where function $f_{\sigma,\sigma'}$ is given by Eq.~(\ref{Eq:f}).
The above results are shown in Fig.~\ref{Fig:3}(a). We plot the ratio $\mathcal{R}(E)$ of spin flip \emph{versus}
spin-conserving probabilities $\mathcal{R}(E)=\bigl[\mathrm{t}(E)_{\sigma,-\sigma}+\mathrm{r}(E)_{\sigma,-\sigma}\bigr]\bigl/\bigl[\mathrm{t}(E)_{\sigma,\sigma}+\mathrm{r}(E)_{\sigma,\sigma}\bigr]$ for different $J/t$. For $J/t = -0.5$ and $-0.05$, i. e.,
when $t^2/U^2\lesssim J/t$, spin-flip transitions are as likely as the spin-conserving ones. For smaller $J/t$, spin-flip probabilities
become proportional to $J^2$, reaching the usual perturbative regime.

Pronounced resonances appear for $U\gg t$. In this limit the singlet and triplet resonant energies are
$E_{\mathrm{res},\ell}\simeq-2\,t^2/U-J(4\ell-3)$,
and $\Gamma \simeq t^3/4U^2$ is the resonance width. The dwell time
$\Delta t_{\mathrm{dw}} = \hbar/\Gamma$ is much greater then the
hopping time $\hbar/t$. We further assume that $\Gamma \lesssim J$,
which is the limit of resonant enhancement of spin relaxation rate. This
condition means that the electron has enough time to precess by the exchange
field before leaking out of the well. The singlet and
triplet resonance peaks are well resolved in this limit.
Equation~(\ref{Eq:t}) now gives Lorentzian,
\begin{equation}
\mathrm{t}(E)_{\sigma,\sigma'}\simeq\sum\limits_{\ell=0,1}\frac{(4\ell\delta_{\sigma, \sigma'}+1)\,t^6\bigl/2U^4}{(E-E_{\mathrm{res},\ell})^2+4t^6/U^4}\label{Eq:r-aprox},
\end{equation}
and similarly Eq.~(\ref{Eq:r}) the reflectivities; $\mathrm{r}_{\sigma,-\sigma} = \mathrm{t}_{\sigma,-\sigma} $, and
$ \mathrm{r}_{\sigma,\sigma} =1 - \mathrm{r}_{\sigma,-\sigma} - \mathrm{t}_{\sigma,\sigma} - \mathrm{t}_{\sigma,-\sigma} $.
Figures~\ref{Fig:3}(b) and \ref{Fig:3}(c) show the comparison of the
 exact and above approximative formulas for $J/t=-0.05$. The peak positions depend on $J$ via $E_{\mathrm{res},\ell}$, but the values at maxima are $J$-independent.
At resonances the spin-flip to spin-conserving probabilities come as $1/3$, see Fig.~\ref{Fig:3}(a):
$25\%$ of scattered electrons change spin. The reason is that a spin up electron forms triplet state $|1,1\rangle$ with 50\% chance,
$|1,0\rangle$ and $|0,0\rangle$ with 25\%. The chance that the electron flips its spin is $50\%$ for each $|1,0\rangle$ and $|0,0\rangle$ states.
This gives the $25\%$ probability for a spin-flip, as we see at resonances.

In \cite{SM} we show, using our 1d model, that an impurity sitting at the barrier site and not inside the well, does not have
such a pronounced effect on the spin-flip probability, justifying our exchange model of hydrogen on graphene that places
$J$ on the hydrogen site only.

In conclusion, we propose that resonant scattering by magnetic impurities in graphene, caused by
vacancies or adatoms, causes the observed fast spin relaxation rates. Resonant enhancement of exchange
interaction, but also of the weaker spin-orbit coupling, opens new prospects for investigating impurity magnetic moments, dynamical polarization of  impurity spins, Kondo physics, and resonant scattering
in graphene.

We thank T.~Wehling for useful discussions, T.~Maas\-sen for providing us the experimental data to Fig.~2,
and P.~Mavropoulos for useful discussions and for pointing to us Ref.~\cite{Mahanti1971:PL}.
This work was supported by the DFG SFB 689 and SPP 1285.

\bibliography{grp_adatom}
\pagestyle{empty}
\begin{widetext}
\includegraphics[width=1.0\columnwidth,angle=0]{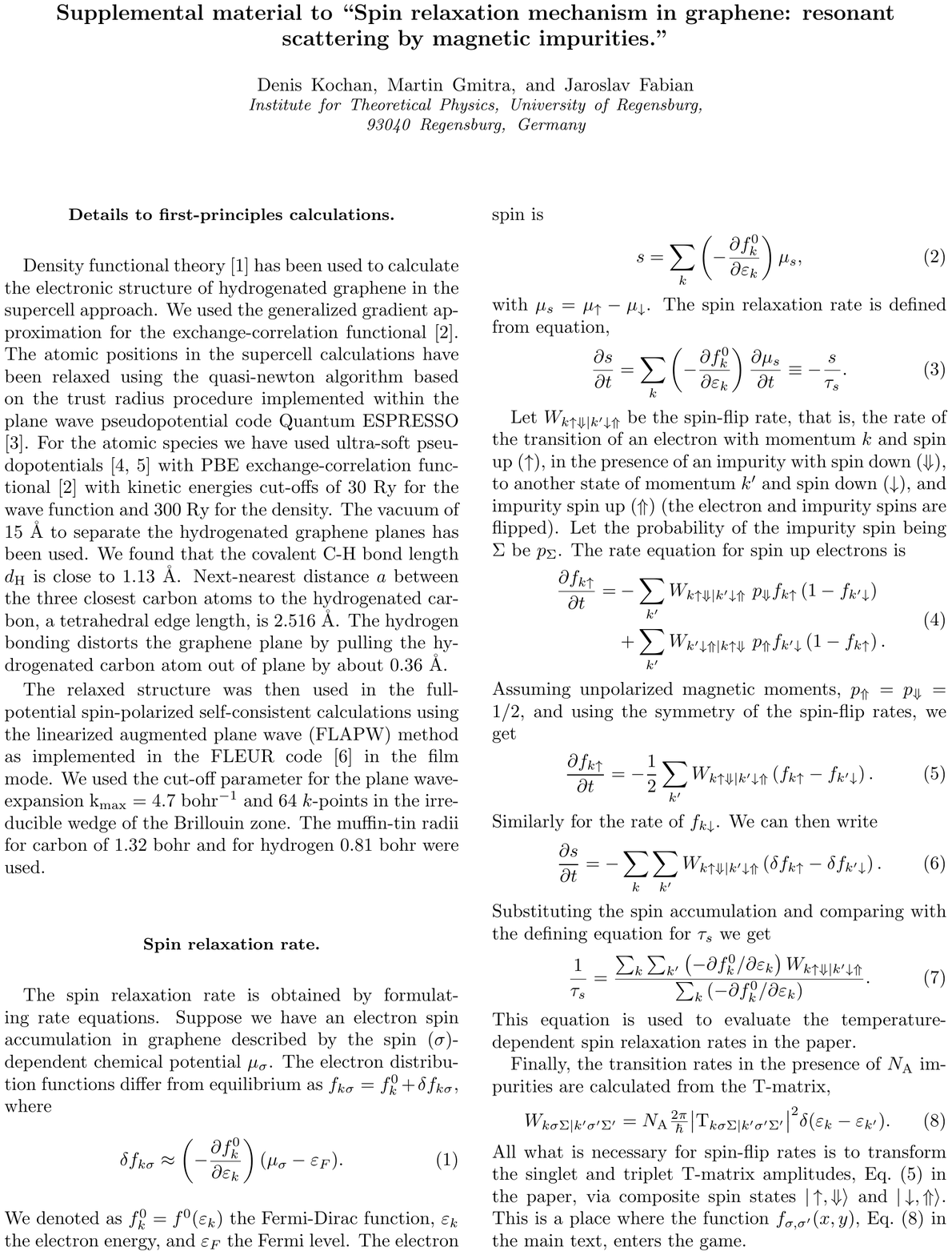}
\includegraphics[width=1.0\columnwidth,angle=0]{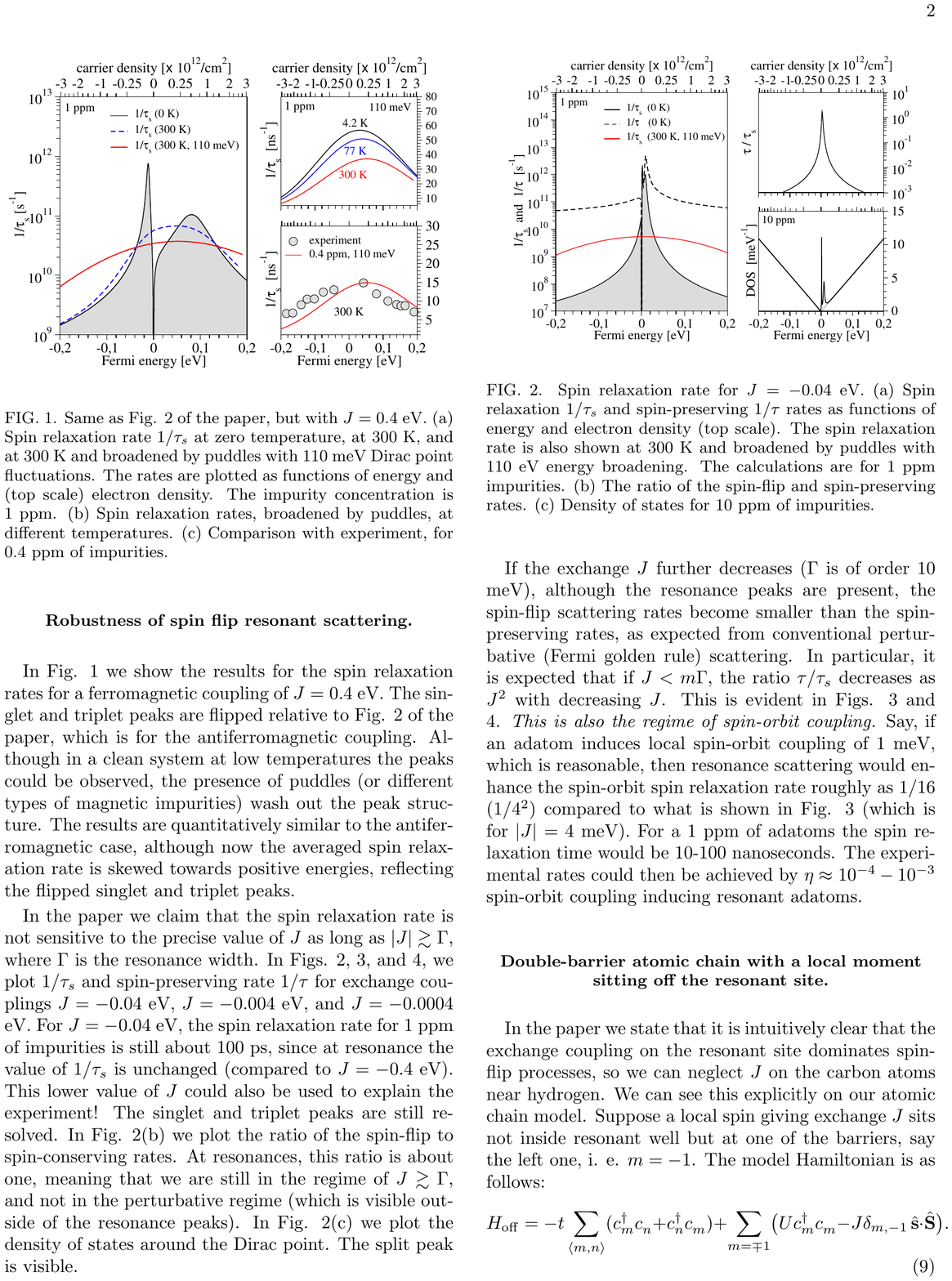}
\includegraphics[width=1.0\columnwidth,angle=0]{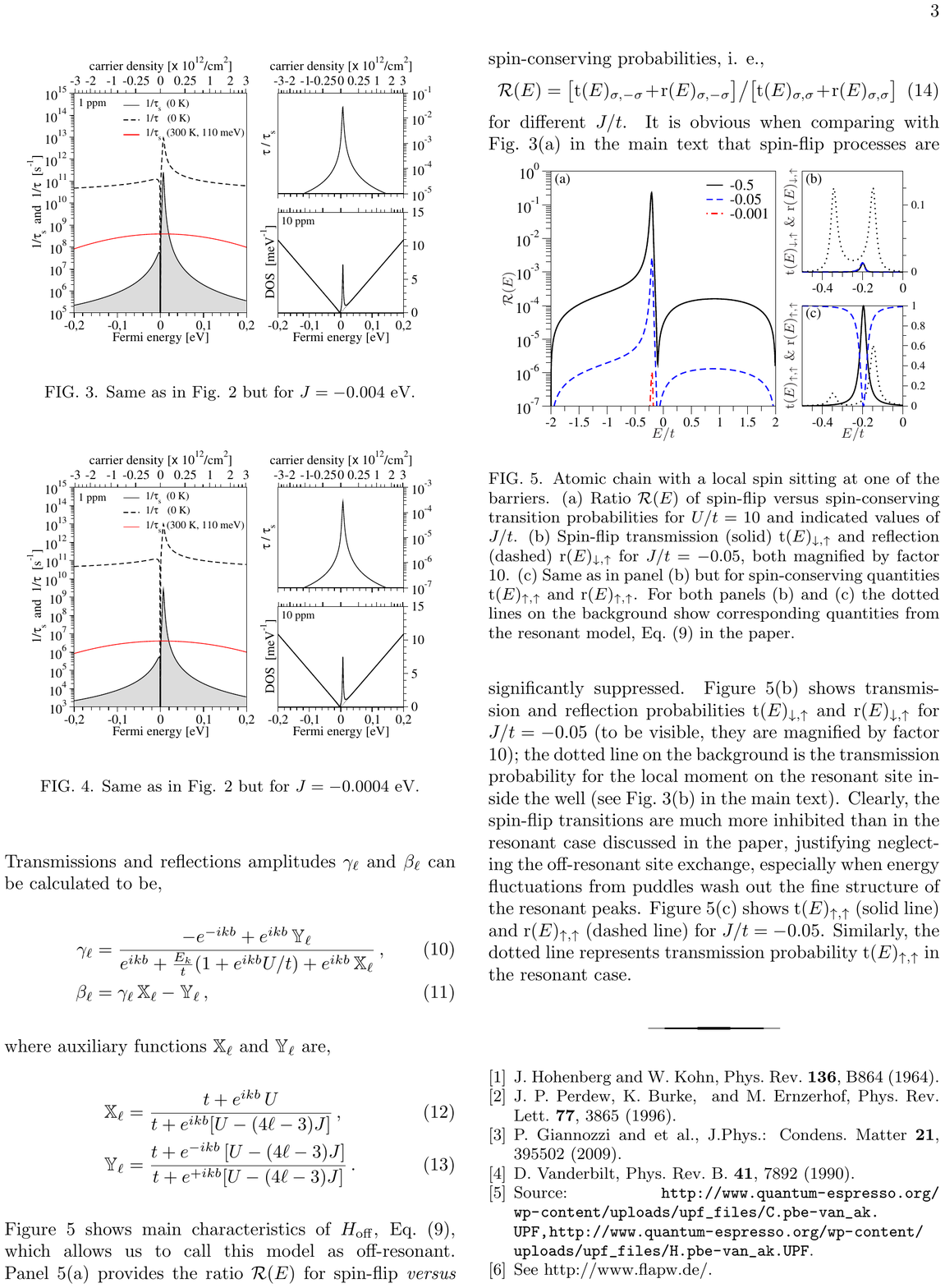}
\end{widetext}

\end{document}